\documentclass[aps,preprint,graphicx]{revtex4}
\usepackage{graphicx}
\usepackage[latin1]{inputenc}

\begin{document}

\title{Nonlinear current-voltage characteristics due to quantum tunneling of phase slips in superconducting Nb nanowire networks}
\author{M. Trezza,$^1$ C. Cirillo,$^1$ P. Sabatino,$^1$ G. Carapella,$^1$ S. L. Prischepa,$^{2}$ and C. Attanasio$^1$}

\affiliation{$^1$CNR-SPIN Salerno and Dipartimento di Fisica \lq\lq E.\,R. Caianiello\rq\rq, Universit\`{a} degli Studi di Salerno, Fisciano (Sa) I-84084, Italy\\
$^2$Belarusian State University of Informatics and
Radioelectronics, P.\,Browka 6, Minsk 220013, Belarus}

\date{\today}

\begin{abstract}

We report on the transport properties of an array of $N \sim 30$ interconnected Nb nanowires, grown by sputtering on
robust porous Si substrates. The analyzed system exhibits a broad
resistive transition in zero magnetic field, $H{\rm \!,}$ and highly nonlinear $V(I)$ characteristics as a function of $H$ which can be both consistently
described by quantum tunneling of phase slips.

\end{abstract}

\pacs{74.78.Na, 73.63.Nm}

\maketitle

Superconductivity in one-dimensional (1D) nanowires has been object of intensive studies
in the last two decades. \cite{ZaikinRep,Herzog96,Vodolazov,Johansson,Bezryadin1,Tian1,Arutyunov1,Altomare}
This great interest rose since superconducting nanowires
involve fundamental phenomena such as macroscopic quantum
tunneling and quantum phase transitions \cite{Giordano,Zimanyi,Zaikin1,Lau} and, in addition, they can find
applications in classical \cite{Bezryadin05} and possibly quantum
information-processing devices \cite{BezryadinQubits,ArutyunovNature} or they can be used as interconnects
in electronic nanostructured devices. \cite{Patel} The main point in the study of superconducting nanowires is to understand how their superconducting properties
change when the 1D limit is approached. Theoretical studies predicted that the resistance of the wire remains different from zero
also at a temperature $T$ well below the critical temperature, $T_{\rm c}$, when the wire diameter is smaller
than its superconducting coherence length. This effect is related to the presence of both thermal activation and
quantum tunneling of phase slips (TAPS and QPS, respectively). \cite{Giordano,Zaikin1,Zaikin2} Several experiments have confirmed the presence of
QPS phenomena in various superconducting materials and systems. \cite{Bezryadin1,Tian1,Lau,Arutyunov3,Cirillo12}

Recently, we developed a nanofabrication approach which can produce samples with physical
properties resembling those of single nanowires. \cite{Cirillo12} The formation of interconnected
networks consisting of ultrathin superconducting Nb nanowires was
achieved by using porous silicon (PS) as a template. Due to the extremely reduced Nb thickness, $d_{\rm Nb} \sim 10$ nm, the deposited material tends to
occupy only the substrate pitch. As a consequence, the sputtered films, at a later stage patterned into microbridges (length $L_{\rm b}=100\ \mu\mathrm{m}
\times{\rm width}\; W_{\rm b}=10 - 20\ \mu\mathrm{m}$) by standard optical lithography and lift-off
procedure, resulted in a network of $250 - 500$ interconnected nanowires, whose average diameter, $\sigma$, was comparable to $\xi$,
so that each individual nanowire behaved as a 1D object. The samples exhibited nonzero resistance over a broad temperature range
below $T_{\rm c}$ and the data were explained considering the occurrence of TAPS and QPS processes. However, no signatures of the presence of QPS phenomena were detected in the
current-voltage characteristics, $V(I)$, of the samples.

The experimental results we show in this letter deal with microbridges [fabricated using Electron Beam Lithography (EBL)] whose in-plane area was 50 times smaller than the previous samples. As a consequence, the resulting network has a drastically lower number of interconnected nanowires $(N \sim 30)$. We examine current transport properties of a sample with $d_{\rm Nb}=12$ nm. As for the data presented in Ref.\,19, the $R(T)$ transition curve is satisfactorily reproduced considering the contribution of quantum fluctuations of the superconducting order parameter. In addition,
$V(I)$ curves, measured at different magnetic fields, have a behavior typical of
1D superconducting individual nanowires which is nicely reproduced considering only QPS processes. \cite{Giordano,ZaikinRep}

\begin{figure}[!ht]
\centering
\includegraphics[height=0.40\textheight]{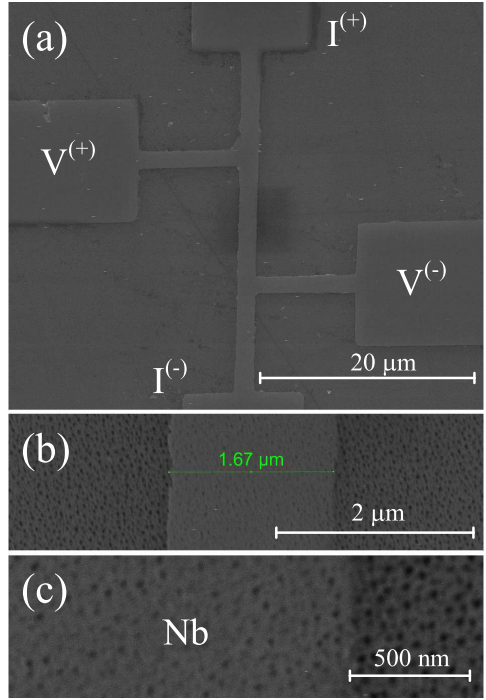}\caption{(Color
online). (a) FE-SEM images of the sample geometry. (b) A middle
portion of the Nb film. (c) Zooming-in of the nanoporous Nb film
edge.} \label{figure1}
\end{figure}

Porous templates were fabricated by electrochemical anodic
etching of n-type, 0.01 $\Omega$ cm, monocrystalline Silicon wafers. \cite{TrezzaJAP08,TrezzaEPL09}
The resulting robust porous substrates, about 2 cm$^{2}$ large, have average pore diameter ${\O}=15$ nm
and average interpore distance $a=50$ nm. The measured samples were patterned in a four-terminal geometry consisting of a pair of
current-carrying Nb electrodes which contact the ($L_{\rm b}=30\ \mu\mathrm{m}
\times W_{\rm b}=1.67\ \mu\mathrm{m}$) nanoporous Nb film
and a pair of voltage pads $10\:\mu\mathrm{m}$ apart, see Fig.\,\ref{figure1}(a). This geometry was realized using EBL, performed in a FEI
Inspect-F field emission scanning electron microscope (FE-SEM)
equipped with a Raith Elphy Plus pattern generator, and a lift-off
procedure. A 1.5-$\mu\mathrm{m}$-thick (both to obtain a clean
lift-off process and to fully cover the substrate pores) positive
tone resist consisting of 9$\%$ polymethylmethacrylate (PMMA) 950K
MW dissolved in anisole was spin-coated onto a nanoporous ${\rm
Si}$ substrate and baked at ${\rm 180\ ^{\circ}C}$ on a
hotplate for 90 sec. The exposition was carried out using a
clearance dose of ${\rm 300\ \mu C/cm^{2}}$ at 30 KV. After
exposure, PMMA was developed in a methyl isobutyl ketone and
isopropyl alcohol solution (1-MIBK:3-IPA) for 30 sec, followed by
rinsing in IPA and deionized water. On this mask Nb ultrathin films were deposited by UHV dc diode magnetron sputtering system with a
base pressure in the low $10^{-8}$ mbar regime following the same
fabrication procedure described elsewhere. \cite{TrezzaJAP08,TrezzaEPL09} Since the effect of the periodicity of the template tends to disappear when $d_{\rm Nb} \gtrsim {\O}$,\cite{TrezzaJAP08}
films thicker than 12 nm can hardly be used to obtain a Nb nanowire network. The value $d_{\rm Nb} \sim 12$ nm
also assures that individual nanowires are continuous and that the presence
of a well defined network is still preserved. \cite{Cirillo12} The final lift-off procedure in acetone allows to obtain the small-area array of interconnected Nb nanowires shown in Figs.\,\ref{figure1}(b) and
\ref{figure1}(c). As already underlined, the effect of the patterning is to reduce the
number $N$ of interconnected wires under study. From the values of $W_{\rm b}$, $L_{\rm b}$, $a$, ${\O}$, and $w=a-{\O}$, $N$ can be easily estimated to be around 30. The effective diameter for each wire is
$\sigma \sim \sqrt{d_{\rm Nb} \cdot w} \sim 20$ nm.
\begin{figure}[!ht]
\centering
\includegraphics[height=0.30\textheight]{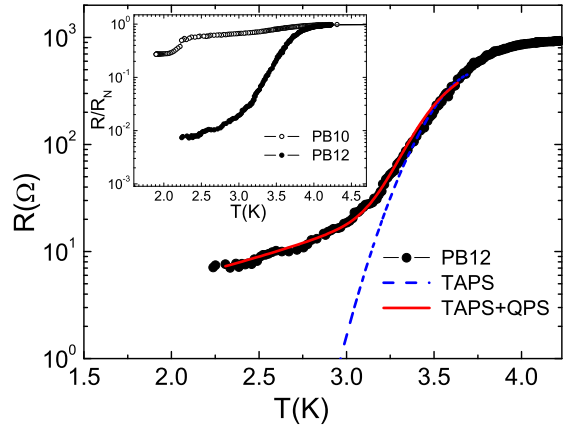}\caption{(Color
online). Zero field transition curve, $R(T)$, of the sample PB12.
Points represent the experimental data while the dashed blue (solid red) line
is the theoretical curve obtained including only TAPS (both TAPS and QPS)
contribution in Eq.\,\ref{eq:Rtot}. The inset shows the $R(T)/R_{\rm N}$ curves of the samples PB10
and PB12.}
\label{RTfit}
\end{figure}

Current transport measurements were resistively
performed in a $^{4}$He cryostat using a standard dc four-probe
technique. The $V(I)$ characteristics were recorded under current
drive condition at different magnetic fields oriented perpendicularly to
the plane of the sample. During the measurement the temperature
stabilization was around 1 mK and, in order to minimize any heating
effect, the sample was kept in contact with liquid
helium. The current biasing was realized by sending
rectangular current pulses to the sample, with the current-on time
being of 12 ms followed by a current-off time of 1 s. Any single
voltage value was acquired at the maximum value of the
current. From resistance versus field, $R(H)$, measurements performed at fixed temperature the perpendicular field phase
diagram, $H_{\rm c2 \perp}(T)$, was obtained and the zero temperature Ginzburg-Landau coherence length was estimated, $\xi(0)=10$ nm.
For further information on both the normal-state and morphological properties of our samples the reader can refer to Ref.\,19, where the central issue of their homogeneity, which can in principle
be responsible of the observed excess of dissipation, \cite{Patel,Lehtinen} has been largely discussed.

Fig.\,\ref{RTfit} shows the $R(T)$ transition curve of the porous Nb
bridge with $d_{\rm Nb}=12$ nm (PB12) measured biasing the sample with a constant current $I_{\rm b}=1$ $\mu$A. The value of the superconducting critical temperature, $T_{\rm c}=3.67$ K, is taken as the temperature at which the
resistance $R$ is reduced to half of $R_{\rm N}=930$ $\Omega$, the low-temperature normal-state
resistance of the nanowire network. Due to the quasi-1D nature of our samples their properties are all extremely sensitive to both $N$ and $d_{\rm Nb}$. To stress this aspect, in the inset of
Fig.\,\ref{RTfit} the resistive curve is reported also for the bridge with $d_{\rm Nb}=10$ nm (PB10) which does not show any transition to the superconducting state down to $T=1.8$ K. Moreover, a nanowire network with the same value for $d_{\rm Nb}$, with a very similar value for $w$ but with $N=250$ had $T_{\rm c}=3.52$ K. \cite{Cirillo12} The main feature of the $R(T)$ curve of the sample PB12 is the long resistance tail at low temperatures which can be associated to the occurrence of 1D superconductivity.
In particular, following the same approach we used in Ref.\,19, the measured temperature dependence of the resistance can be reproduced using \cite{Lau}

\begin{equation}
R(T) = [R_{\rm N}^{-1} + (R_{\rm TAPS}+R_{\rm QPS})^{-1}]^{-1} \label{eq:Rtot}
\end{equation}
where $R_{\rm TAPS}$ and $R_{\rm QPS}$ represent the contribution to the resistance due to thermal activation and quantum tunneling of phase slips, respectively. It is
$R_{\rm TAPS} \sim \sqrt{F(T)/k_{\rm B}T}\exp(-F(T)/k_{\rm B}T)$ \cite{Zaikin2,Cirillo12} and, with exponential accuracy, \cite{Zaikin1,Cirillo12,Bae,Lehtinen}
\begin{equation}
R_{\rm QPS}(T) \approx
A\frac{R_{\rm Q}^{2}}{R_{\rm N}}\frac{L^{2}}{\xi^{2}(0)}\exp\Bigg[-A\frac{R_{\rm Q}}{R_{\rm N}}\frac{L}{\xi(T)}\Bigg]
\label{eq:QPS}
\end{equation}
where the quantum resistance $R_{\rm Q}\approx 6.45$ k$\Omega$, $L$ is the nanowire length and $A$ is a numerical parameter of the order of the unity.
$\xi(T)=\xi(0)/\sqrt{1-T/T_{\rm c}}$ and $F(T)=F(0)(1-T/T_{\rm c})^{3/2}$
are the temperature dependent coherence length and phase slip
activation energy, respectively. The results of the theoretical analysis are reported in
Fig.\,\ref{RTfit} where the best fit curve, which contains $T_{\rm c}$, $A$, and $L$ as adjustable parameters, and nicely reproduces the experimental data, is represented by the solid red line. Interestingly, the fitting procedure gives $T_{\rm c}=3.64$ K, a value
very close to the measured one, and $A \approx 0.1$ a number which is in good agreement with the values reported for this parameter in the case of single Ti nanowires. \cite{Lehtinen} Moreover, the length $L$ of the single nanowire appearing in the expression of
$R_{\rm QPS}$  was treated as a fitting parameter. \cite{Cirillo12} The obtained value, $L \approx 70$ nm, appears to be rather small if directly compared to the estimated length, around $10\ \mu\mathrm{m}$, of a single
continuous nanowire present in the network. This discrepancy could be explained considering that $L$ is not a quantity which can be strictly defined for the network. Moreover, the
theoretical models we used were developed for the case of an individual wire and in a disordered network $L$ is more likely a \lq\lq percolation\rq\rq \,\,length. If, on the other hand,
only the TAPS dependence is considered and the QPS term is disregarded in Eq.\,(\ref{eq:Rtot}), the obtained curve
strongly deviates from the measurements (dashed blue line in Fig.\,\ref{RTfit}).

\indent
\begin{figure}[!ht]
\centering
\includegraphics[height=0.30\textheight]{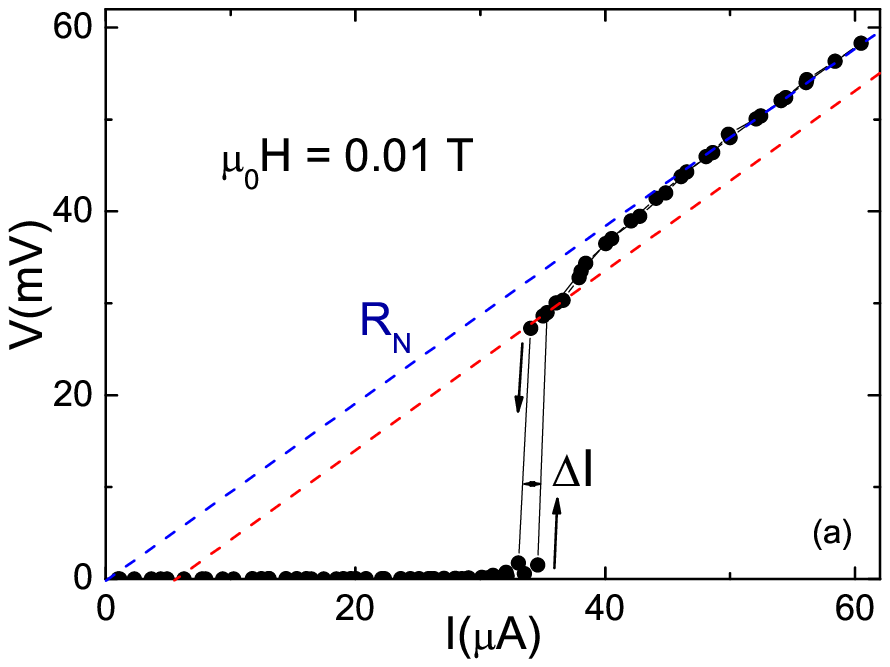}
\includegraphics[height=0.30\textheight]{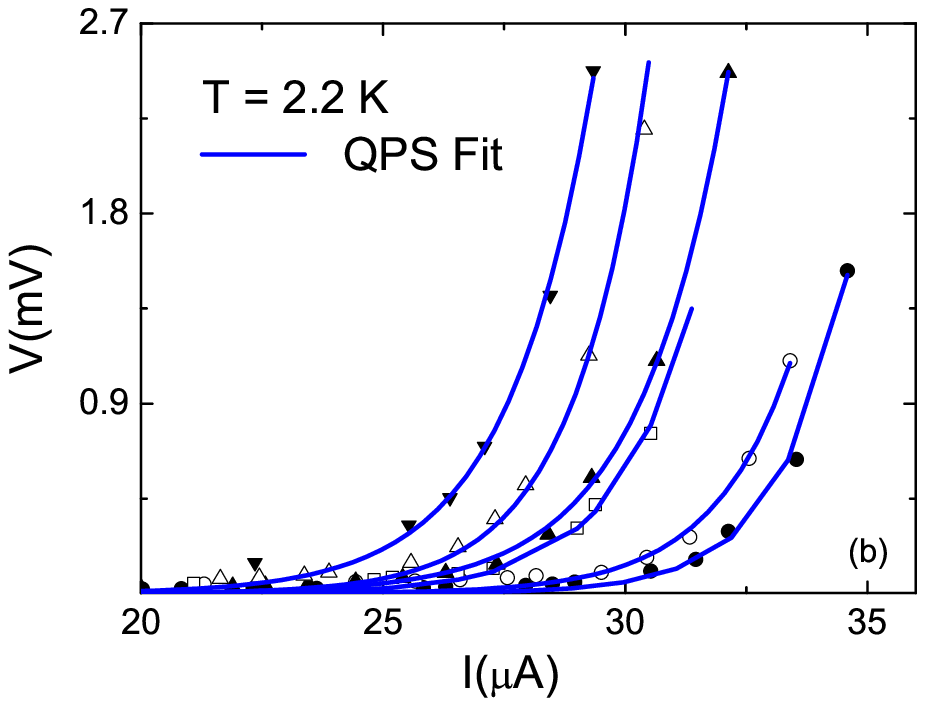}\caption{(Color
online).  (a) $V(I)$ dependence measured at $T=2.2$ K and $\mu_{\rm 0} H=0.01$ T. The red dotted line shows the resistance due to the single phase slip
entering the sample. The blue dotted line indicates the normal-state resistance, $R_{\rm N}$. (b) Nonlinear voltage versus current characteristics measured at $T = 2.2$ K
at different fields. Applied
magnetic fields are, from right to left, 0.01, 0.025, 0.05, 0.06, 0.08
and 0.12 T. Blue lines are the QPS fits to the data.} \label{IVFit}
\end{figure}
To further highlight the role played at low temperature by QPS processes we measured $V(I)$ characteristics as a function of the field \cite{Xiong} at the lowest temperature reached when the resistive transition measurement was recorded. The curves were obtained, in pulse mode, sweeping the current first upward and then downward in presence of several magnetic fields. From the $V(I)$ curves we have preliminarily estimated the critical current of the single wire, $i_{\rm c}$, through the relation $I_{\rm c}=(N+1)i_{\rm c}$, where $I_{\rm c}$ is the critical current of the entire network. \cite{vanderZant} $i_{\rm c}$ is also related to the
single wire depairing current, $i_{\rm dp}$, via $i_{\rm c}=2i_{\rm dp}/3 \sqrt{3}$. \cite{vanderZant} At zero field and $t \simeq 0.6$ we have $I_{\rm c} = 32 \,\,\mu$A so that $i_{\rm c} \sim 1 \,\,\mu$A. This number is only a factor of two lower that the one obtained at zero field and $t \simeq 0.5$ on the network of 250 nanowires with a similar value of $\sigma$ where $I_{\rm c} = 470 \,\,\mu$A. \cite{Cirillo12} As a consequence, also the zero temperature depairing current density \cite{KL} of the single wire $j_{\rm dp}(0)=j_{\rm dp}(t)(1-t)^{-1.5} \sim [i_{\rm dp}(t)(1-t)^{-1.5}]/\sigma^2 \simeq 1.9 \times 10^{10}$ A/m$^2$ is comparable to the one reported in the case of the network with $N=250$ and in the case of perforated Nb films. \cite{Cirillo12,Sabatino} This result indicates that the further reduction of $N$ does not significantly depress the superconducting properties of the single wire and confirms, again, the good quality of the sample ruling out the presence of tunneling barriers at the grain boundaries. \cite{Patel} In Fig.\,\ref{IVFit}(a) we show the $V(I)$ curve measured at $T=2.2$ K $(t \equiv T/T_{\rm c} \simeq 0.6)$ and $\mu_{\rm 0} H = 0.01$ T. As expected, being the temperature far from $T_{\rm c}$, the curve, as can be inferred from the red dotted line, does not show multiple steps corresponding to more and more phase slip lines entering the sample but only one large step is observed. \cite{Bezryadin5} It is also important to notice that, in small applied fields, a clear hysteresis was present, which became weaker when the field was increased, completely disappearing for $\mu_0 H > 0.14$ T. We rule out the heating as possible cause of the hysteresis since, after the phase slip has entered the sample, all the $V(I)$ curves, before reaching the normal state, are linear and they do not extrapolate to zero. \cite{Bezryadin5} In particular, the width of the hysteresis ranged from $\Delta(I) \equiv I_{\rm up}(10 \,\,{\rm mV}) - I_{\rm down}(10 \,\,\rm{mV}) =1.50$ $\mu$A for $\mu_0 H = 0.01$ T
to the value $\Delta(I)=0.40$  $\mu$A for $\mu_0 H = 0.14$ T. This behavior is qualitatively similar to what observed in NbSe$_2$ nanowires where the hysteresis vanished as the temperature was raised towards $T_{\rm c}$ \cite{Hor} though, in our case, this value of the field is considerably far from $\mu_0 H_{\rm c2 \perp}(t \simeq 0.6) \sim 1.2$ T.

$V(I)$ characteristics measured at $T=2.2$ K for several fields are shown in Fig.\,\ref{IVFit}(b) where, for the sake of legibility, only the curves recorded
sweeping the current upward are reported. The curves have been fitted using separately the two exponential
expressions valid if TAPS or QPS are responsible for the phase slippage. For thermal activation of phase slips it is \cite{ZaikinRep,Zaikin2}
\begin{equation}
V_{\rm TAPS} = {2 \pi h \over e} \gamma_{\rm TAPS} \sinh\Bigg({h I \over 4 e k_{\rm B} T}\Bigg)
\label{eq:VTAPS}
\end{equation}
with $\gamma_{\rm TAPS} \sim (R_{\rm TAPS}/R_{\rm Q}) (k_{\rm B} T/h)$.\cite{Zaikin2} If quantum tunneling of phase slips takes place $V_{\rm QPS}$ has a complicated expression which essentially gives for the $V(I)$ dependence a $\sim \sinh(I)$ behavior. \cite{ZaikinRep,Zimanyi} However, in the high-current (low-temperature) limit it simply reduces to \cite{ZaikinRep,Arutyunov1,Zimanyi}
\begin{equation}
V_{\rm QPS} \sim I^{2 \mu -1}
\label{eq:VQPS}
\end{equation}
with $\mu = R_{\rm Q}/R_{\rm qp}$. Here $R_{\rm qp}$ is a resistance of the order of $R_{\rm N}$ \cite{Arutyunov1} which can be considered as a fitting parameter. The best fits to the $V(I)$ curves, shown by solid blue lines in Fig.\,\ref{IVFit}(b), were indeed obtained using Eq.\,(\ref{eq:VQPS}). The values extracted for the parameter $R_{\rm qp}$ depend on the magnetic field and are all lower than $R_{\rm N}$ going from $\sim 600$ $\Omega$ at $\mu_0 H=0.01$ T to $\sim 800$ $\Omega$ at $\mu_0 H=0.14$ T. This result is consistent with the fact that, since $R_{\rm qp}$ represents the low-temperature residual resistance, \cite{Arutyunov1} it should tend to $R_{\rm N}$ when the applied magnetic field becomes larger. When the field is further increased beyond 0.14 T the best fit curves do not satisfactorily reproduce the experimental data. Surprisingly, this field is the same at which the hysteresis in the $V(I)$ curves vanishes. We do not have any clear explanation for this coincidence. If, on the other hand, we try to interpret the $V(I)$ characteristics within the model of thermal activation of phase slips, coherently with the results established from the analysis of the $R(T)$ curve, experimental data cannot satisfactorily be fitted if any plausible set of parameters entering $\gamma_{\rm TAPS}$ is considered in the Eq.\,(\ref{eq:VTAPS}). This result strengthens the idea that current transport properties of the superconducting nanowire network at the chosen (lowest) temperature are mainly due to quantum tunneling of phase slips. Furthermore, it is also worth mentioning that in the case of the $V(I)$ characteristics measured for the wider network \cite{Cirillo12} it was not possible to fit the curves, in conjunction with the resistive transition, using the same procedure adopted above.

In conclusion, quantum fluctuations of the superconducting order parameter were consistently revealed from both $R(T)$ and
$V(I)$ measurements in superconducting Nb nanowire networks
patterned on PS templates. The reduction of the number of interconnected wires down to $N \sim 30$ made QPS phenomenon the dominant contribution to the current transport properties of the system. All the data were coherently reproduced in the framework of theoretical models elaborated to describe QPS processes. The analyzed system, obtained starting from a robust and macroscopically large substrate, reveals fascinating quantum effects and shows high values of the critical current density. This last occurrence makes the system of potential use as 1D interconnection in complex nanodevices. \cite{Hor}

M. Trezza and P. Sabatino acknowledge financial support from
\lq\lq PON Ricerca e Competitivit\`{a} 2007-2013\rq\rq\,  under grant
agreement PON NAFASSY, PONa$3\_\,00007$.

\end{document}